# Adaptive real-time dual-comb spectroscopy

Takuro Ideguchi[1*], Antonin Poisson[2*], Guy Guelachvili[2], Nathalie Picqué[1,2,3 †], Theodor W. Hänsch[1,3]

1. Max Planck Institut für Quantenoptik, Hans-Kopfermann-Str. 1, 85748 Garching, Germany
2. Institut des Sciences Moléculaires d'Orsay, CNRS, Bâtiment 350, Université Paris-Sud, 91405 Orsay, France
3. Ludwig-Maximilians-Universität München, Fakultät für Physik, Schellingstrasse 4/III, 80799 München, Germany

* These authors contributed equally to this work
† Corresponding author: nathalie.picque@mpq.mpg.de

**Abstract**

*With the advent of laser frequency combs [1,2], coherent light sources that offer equally-spaced sharp lines over a broad spectral bandwidth have become available. One decade after revolutionizing optical frequency metrology, frequency combs hold much promise for significant advances in a growing number of applications including molecular spectroscopy [3-18]. Despite its intriguing potential for the measurement of molecular spectra spanning tens of nanometers within tens of microseconds at Doppler-limited resolution, the development of dual-comb spectroscopy [8-18] is hindered by the extremely demanding high-bandwidth servo-control conditions of the laser combs. In this letter, we overcome this difficulty. We experimentally demonstrate a straightforward concept of real-time dual-comb spectroscopy, which only uses free-running mode-locked lasers without any phase-lock electronics, a posteriori data-processing, or the need for expertise in frequency metrology. The resulting simplicity and versatility of our new technique of adaptive dual-comb spectroscopy offer a powerful transdisciplinary instrument that may spark off new discoveries in molecular sciences.*

Fourier transform spectroscopy [19,20] is the mature solution to a variety of problems [21-26] from the research laboratory to the manufacturing floor. The Michelson-based Fourier transform interferometer is known for more than 40 years as a benchtop instrument of great value in analytical sciences because it measures a broad range of optical frequencies simultaneously with accuracy and sensitivity. Recently, the precisely spaced spectral lines of a laser frequency comb have been harnessed for new techniques of dual-comb Fourier transform spectroscopy [8-16]. Due to interference between pairs of optical comb lines, the optical spectrum is effectively mapped into the radio frequency region, where it becomes accessible to fast digital signal processing. Dual-comb spectroscopy holds much promise for outperforming Michelson-based Fourier transform spectroscopy: the recording time and resolution may be improved a million-fold, while keeping the ability for broad spectral span in any spectral region. However, the technique has not realized its full potential yet, mostly due to the difficulty of synchronizing the pulse trains of two combs within interferometric precision. In this letter, we present an easy-to-implement technique of real-time dual-comb spectroscopy that only requires free-running femtosecond oscillators. Long measurement times, complex locking electronics and time-consuming computer algorithms are not needed anymore.

Dual-comb spectroscopy can be considered with the metaphor of a sampling oscilloscope. Pulses from comb 1 excite molecular free induction decay and pulses from comb 2 sample the waveform of this decay interferometrically. For simplicity, we at first ignore carrier-envelope phase shifts. At best, interferometric samples are taken at intervals $1/f$ (e.g. $10^{-8}$ s) where $f$ is the repetition frequency of comb 2 (e.g. 100 MHz). If $\Delta f$ is the difference of comb repetition frequencies (e.g. 100 Hz), consecutive interferometric samples result from pulse pairs showing a time separation increased by an amount $\Delta\tau = \Delta f/f^2$ (e.g. $10^{-14}$ s). The "sampling oscilloscope" effectively stretches the waveform of the free induction decay signal by a factor $s = f/\Delta f$ (e.g. $10^6$). Signal frequencies in this waveform are transformed down from the optical to the radio-frequency region by the same factor s. In the experiment, the signal of the photodetector is electronically low-pass filtered in order to suppress the pulse repetition frequency $f$. The time-stretched waveform appears thus as a continuous electronic signal. This signal is digitized at a constant clock rate determined by the data acquisition board of the computer. For ideally stable frequency combs, the waveform thus sampled can be Fourier-transformed to reveal the signal spectrum. In the real world, major difficulties arise from the residual instabilities of the frequency combs, even when these benefit from state-of-art stabilization. The time intervals between excitation and sampling pulses are indeed subject to some variations δt, which appear in the detector signal stretched in time by the factor s. Let us now consider the slippage of the carrier phase relative to the pulse envelope due to laser dispersion. If the carrier-envelope slippage frequencies of the two combs differ by $f_{ce}$, the relative phases of pump and probe pulse will change by an additional $2\pi f_{ce}/f$ between two interferometric samples, and all frequencies in the Fourier spectrum of the detector signal will be translated by $f_{ce}$. As long as $f_{ce}$ is constant, the spectral translation can be accounted for if $f_{ce}$ is measured or if the absolute frequency of some spectral feature is known.

To record a distortion-free real-time interferogram, the timing jitter δt between subsequent pulses must be kept lower than 10 attoseconds. Otherwise, chromatic artifacts, called phase errors, spoil the spectrum [27] and cannot be accounted for *a posteriori*. Stabilizing the combs against state-of-the-art cavity-stabilized continuous-wave lasers with a hertz-level line-width, far exceeding the stability available from common microwave oscillators, enables to record satisfactory interferograms averaged over several seconds [10,13,14]. Possible phase errors in ms-time scale acquisitions may be thus cleaned up. However this arduous approach is hardly implementable outside the best frequency metrology laboratories and wastes the sub-ms recording time advantage. Alternatively, simultaneously recording the interferometric signal and the fluctuations of two loosely-stabilized combs against two continuous-wave lasers to compute *a posteriori* corrections has been demonstrated [15]. This solution, with multiple data acquisition channels and additional computational time, also prevents fast acquisition rates. Actually dual-comb interferograms recorded in times shorter than a millisecond are spoiled by intrinsic comb noise [28] that is challenging to eliminate. This contrasts with the less demanding requirements on the frequency comb stability for frequency metrology, where the beat note between a comb line and a continuous-wave laser is integrated over several seconds. As a consequence, an implementation of real-time Doppler-limited dual-comb spectroscopy allowing distortion-free spectra to be recorded has not been reported so far.

In this letter, we describe a simple technique, which makes it possible to record real-time high quality molecular spectra with free-running femtosecond lasers, and we discuss its experimental performance with commercial fiber lasers emitting around 1.55 μm. Dramatic compensation for the timing and carrier-envelope phase shift variations is achieved by correcting the interferometric signal and triggering the data acquisition with two adaptive clock signals. Such signals, which automatically provide the needed correction, can be derived from radio-frequency beat notes between lines from the two different combs.

Two different radio-frequency beat signals between different individual lines of the two combs with beat frequencies $f_a$ and $f_b$ are used to produce two electronic signals that can both account for timing jitter and variations of the relative carrier envelope phase. A filtered beat note signal with a frequency $f_a$ can be considered as a time-stretched sampled waveform of an optical comb line. Frequency combs can produce random variations δϕ of the phase difference between pump and probe pulse. These phase variations δϕ are directly imprinted onto each frequency component of the detected signal. Electronic multiplication of the signal at $f_a$ with the sampling detector signal can then compensate for variations of the relative carrier envelope phases for all optical frequencies but only partly for timing changes. Any timing variation δt will shift the beat signal by the stretched amount s δt. If the pulses were only subject to such timing variations, all frequencies in the spectrum would benefit equally from the produced correction signal. The difference frequency between the beat signals $f_a$ and $f_b$ is immune to common phase variations δϕ and can provide the second required adaptive clock signal, which just compensates for timing variations. This second adaptive signal is used to trigger data acquisition.

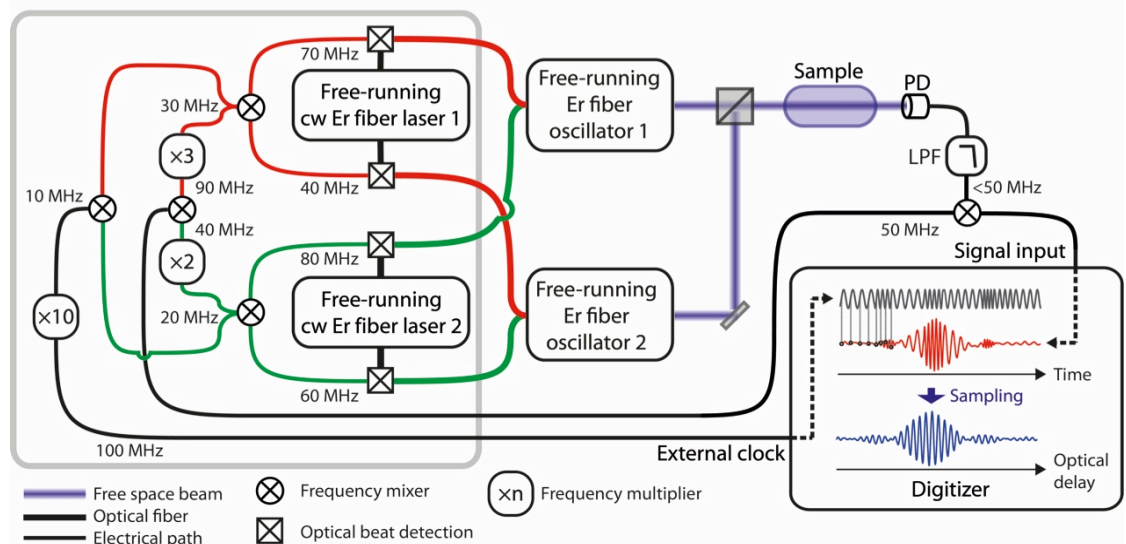

**Figure 1: Sketch of adaptive dual-comb spectroscopy.**
The output of a free-running femtosecond (fs) Er-doped fiber laser is combined with that of a second free-running fs Er-doped fiber laser with a slightly different repetition frequency. Both beams probe the sample. Their temporal interference pattern is recorded with a fast photodetector PD, filtered, mixed with the first adaptive signal and digitized. Each fs laser simultaneously beats with two independent narrow line-width continuous-wave free-running erbium fiber lasers to isolate single longitudinal modes of each fs laser. Each pair of beating signals with a given cw laser is electronically mixed in order to produce an electric signal reporting the relative fluctuations between two modes of the two combs. These signals are frequency-multiplied and mixed according to a manner described in the text. They produce two adaptive signals. The first adaptive signal is mixed with the interferometric signal to compensate mainly for the phase variations while the second adaptive signal serves as the external clock signal for data acquisition and accounts for the timing fluctuations only.

In an experimental implementation of the scheme (Fig.1), two erbium free-running commercial femtosecond lasers emitting in the 1550 nm region are used. Their repetition frequencies are 100 MHz and differ by about 350 Hz. The experimental set-up is placed in a basic laboratory environment, without air-conditioning system, vibration isolation, or dust protection. Each free-running femtosecond erbium-doped fiber laser has two fibered output ports.

For each of the femtosecond lasers, the port that emits an average power of about 20 mW is used for the dual-comb interferometric set-up. The two beams are combined to interrogate the sample and beat on a fast InGaAs photodiode. The time-domain interferometric signal exhibits a periodic succession of huge bursts every $1/\Delta f$ when femtosecond pulses from the two lasers coincide. A low-pass filter essentially suppresses the unmodulated part of the interferogram. This part consists of pulses generated at a repetition frequency of 100 MHz by the two lasers. In order to synchronize the interferometric signal with the adaptive signals, one has to compensate for the delays induced by the different optical paths and the various electronic components. An electronic delay line is therefore inserted in the interferometric signal path. The interferometric electric signal is mixed with the first adaptive signal described below. A low-noise 100-MHz-bandwidth amplifier with variable gain scales the filtered mixed signal to the full range of a 14-bit analog-to-digital converter. This high-resolution digitizer, onboard a personal computer, has 65-MHz-bandwidth and sampling rate capabilities of up to 125 MSamples/s. It is externally triggered with the second adaptive clock signal described below.

The second output port of the femtosecond lasers serves to generate the adaptive signals. Each laser independently beats with two free-running continuous-wave erbium-doped fiber laser that emit at 1534 nm and 1557 nm, respectively. By beating each comb with a continuous-wave laser emitting at a frequency $f_{cw}$, two beat notes $f_{cw}$-$f_{1,n}$ and $f_{cw}$-$f_{2,n}$ are produced, thus isolating one line of each of the two femtosecond lasers is isolated. The electric signals are filtered and a delay line adjusts the timing between the two beat signals. The two beat notes are electronically mixed so that the contribution of the continuous-wave laser cancels to generate the signal $f_a$ =$f_{1,n}$-$f_{2,n}$. In this manner, two beat notes between individual comb lines from the two combs are produced: $f_a$=$f_{1,n}$-$f_{2,n}$ and $f_b$=$f_{1,m}$-$f_{2,m}$. In our experiments, the frequency of these beat notes is about 30 and 20 MHz, respectively.

As the radio-frequency spectrum spans the free spectral range $0 - f/2$, i.e 0-50 MHz, the signals at frequencies $f_a$ and $f_a$-$f_b$ are not high enough to avoid aliasing. The first adaptive signal is fabricated from a signal at frequency $3f_a$-$2f_b$ produced by a combination of frequency multipliers and mixers. The subsequent 50 MHz signal is mixed with the interferometric signal before digitalization by the data acquisition board. The second adaptive signal results from the frequency-multiplication by ten of the beat signal at frequency $f_a$-$f_b$. A 100 MHz signal thus provides the adaptive clock signal that is connected to the external clock input of the digitizing data acquisition board. We note that combs where the carrier-envelope offset cancels, producible by difference frequency generation [16,29], would only require one adaptive signal to be produced.

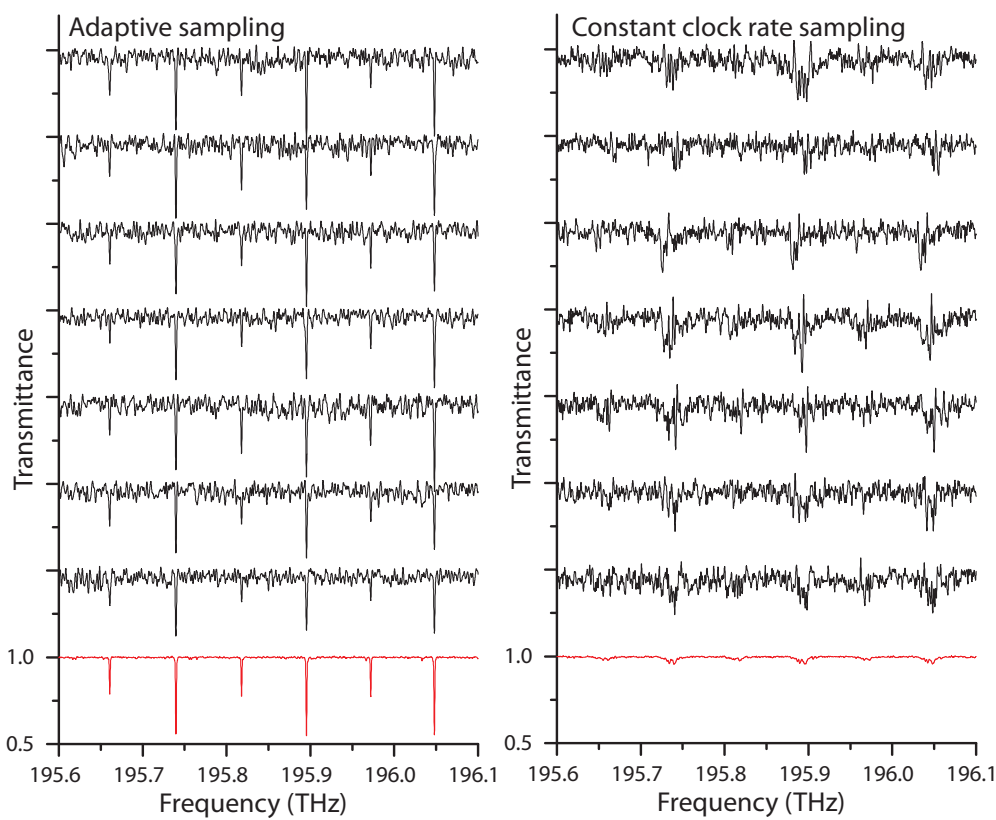

**Figure 2. Portions of experimental and computed spectra of acetylene.**
(a) Zoom of a span of 0.5 THz of seven consecutive adaptive individual spectra (in black) and the average of 200 such spectra (in red). Each spectrum is measured over a 14.5 THz span within 467 µs.
(b) Spectra digitized at the constant clock rate of the data acquisition board and identical conditions than in (a) otherwise. The acetylene line profiles are strongly distorted and the drifts in the frequency scale do not allow for averaging.

The absorption spectrum by a single pass gas cell filled with acetylene in natural abundance is recorded in the region of the $\nu_1+\nu_3$ combination band. Fig.2 displays a 0.5 THz zoomed part of a 14.5 THz spectral span. Seven consecutive spectra are displayed. Each is measured within 467 µs with 1.1 GHz apodized resolution at a refresh rate of 350 Hz. The reproducibility of consecutive adaptive spectra permits efficient averaging whereas free-running lasers without the adaptive sampling scheme produce strongly distorted spectra and do not allow for dual-comb spectroscopic measurements. To make the excellent quality of the experimental adaptive spectra obvious, we have increased (Fig. 3, Fig. 4) the number of averaging to 200 and compared the spectrum with a spectrum computed from the HITRAN database [30]. The good signal-to-noise ratio in the experimental adaptive spectrum clearly reveals the rotational lines of the $\nu_1+\nu_3$ cold band, the $\nu_1+\nu_3+\nu_4^1-\nu_4^1$ and $\nu_1+\nu_3+\nu_5^1-\nu_5^1$ hot bands of $^{12}C_2H_2$ and even of the $\nu_1+\nu_3$ cold band of $^{12}C^{13}CH_2$ though its relative concentration is decreased about a hundred fold. The level of the residuals between the adaptive spectrum and the calculated one remains below 1.5 %. Calibration against two molecular lines present in the adaptive spectrum sets the frequency accuracy on the other line positions to 10 MHz (relative accuracy: 5 10$^{-8}$) at 1 ms measurement time. The very good agreement between experimental and computed profiles shows the suitability of our technique for line intensities or concentration measurements. An equivalent way to assess the efficiency of our adaptive scheme is, rather than averaging the consecutive spectra, to Fourier-transform a long time-domain sequence thus allowing for resolving the comb lines and showing that the coherence between the two combs is maintained during the measurement time. Fig. 4 shows a spectrum of acetylene spanning over 12 THz recorded in 2.7 s with 268 $10^6$ samples. The 120 000 individual comb lines are resolved across the full spectral span and their apodized instrumental linewidth is 202 kHz in the optical domain (0.7 Hz in the radio-frequency domain). In comparison, a Michelson-based Fourier transform spectrometer would record a similar spectral domain, with $10^5$ spectral elements and a resolution at best of the order of 100 MHz, in several hours. The quality of our present results also far exceeds what is obtained with the commercially available servo-controlled comb systems referenced to a radio-frequency clock like a GPS or a hydrogen maser, which do not allow for resolving the comb lines.

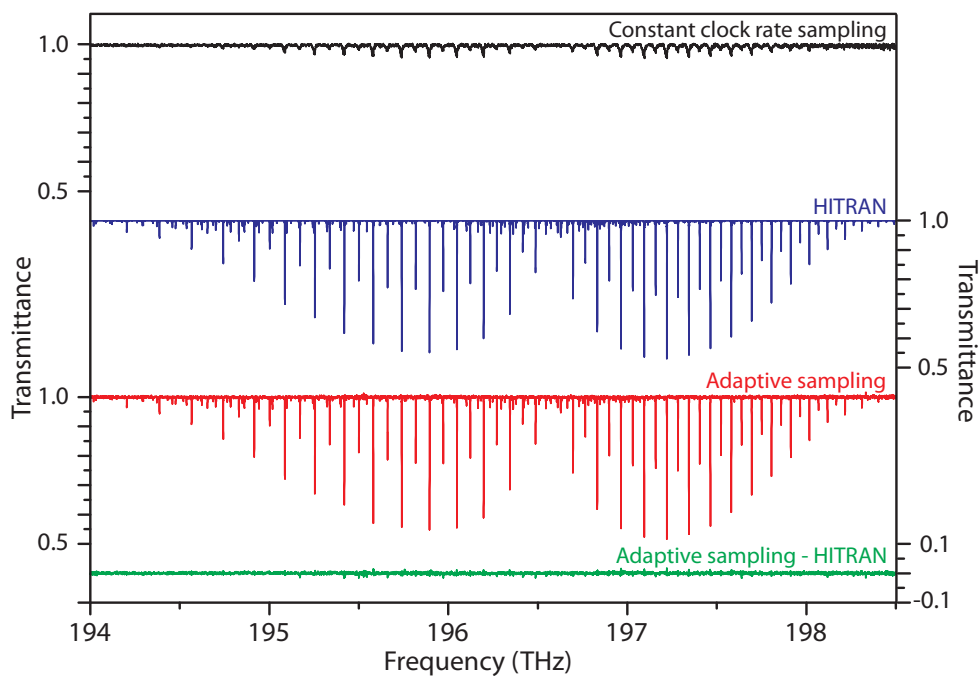

**Figure 3. Illustration of the quality of the adaptive sampling spectra.**
Experimental absorption spectra of $^{12}C_2H_2$ and $^{12}C^{13}CH_2$ sampled at adaptive and constant clock rates, respectively, are compared to a spectrum computed from the line parameters available in the HITRAN database [28]. The cell is filled with 1.6 Torr of acetylene. Both experimental spectra are measured with free-running fs lasers in identical conditions except for the data acquisition clock. The difference in repetition frequencies between the two free-running fs lasers is set to 344.9 Hz. Two hundred consecutive spectra with an apodized resolution of 1.1 GHz, each measured within 467 µs, are averaged, resulting in a total measurement time of 93 ms and a total experimental time of 580 ms. The agreement of the experimental adaptive spectrum with the computed spectrum from the $^{12}C_2H_2$ and $H^{12}C^{13}CH$ line parameters of the HITRAN database confirms the suitability of our technique for measurements with Doppler-limited resolution. When sampled at the constant clock rate of the digitizing data acquisition board, the spectrum is strongly distorted due to the relative fluctuations of the repetition frequency and carrier envelope phase slips of the fs lasers. This lowers the sensitivity and scrambles the spectral features.

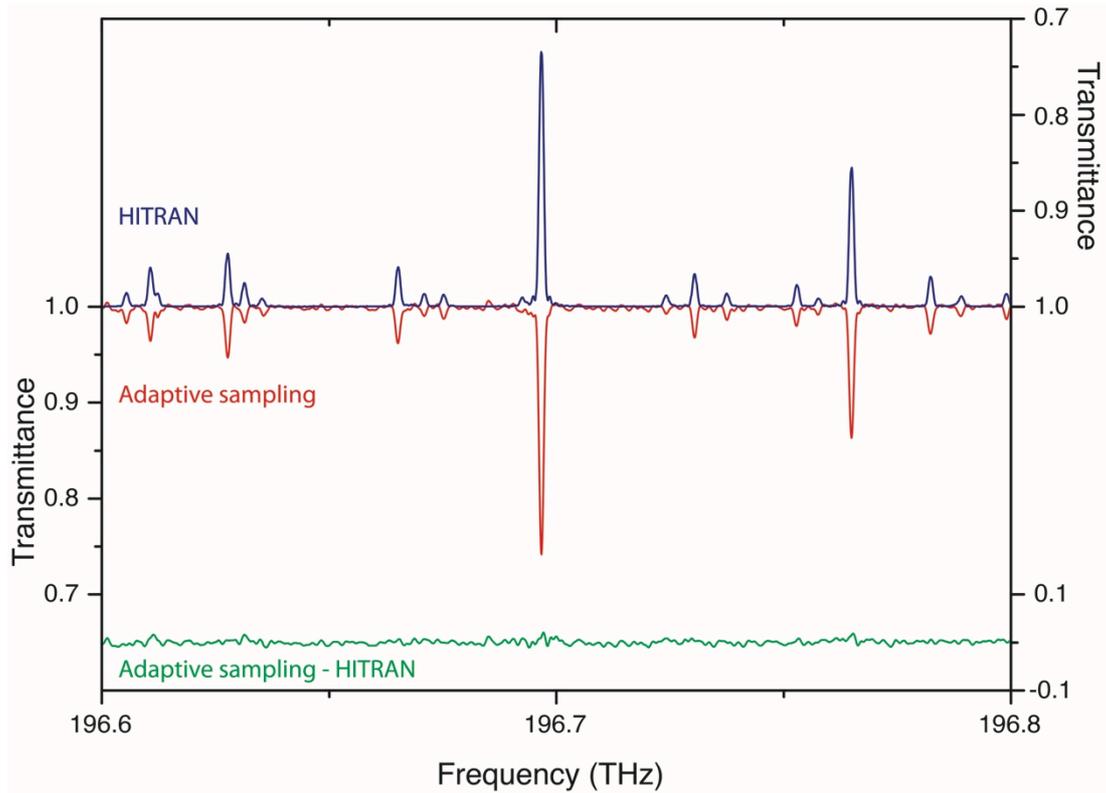

**Figure 4. Zoom of a 0.2 THz portion of the spectra shown in Figure 3.**
The adaptive experimental spectrum and the spectrum computed from the HITRAN data shown in Figure 3, together with the residuals, are expanded both in the x- and y-scales to give better appreciate the high resolution details.

Our adaptive scheme perfectly compensates for the fluctuations of the femtosecond lasers at any optical frequency. In this demonstration, all lasers are free-running and high-quality molecular spectra down to Doppler-limited resolution are recorded under such conditions. Therefore, the instrument is suitable for optical diagnostics in a variety of applications. If absolute direct frequency calibration via the frequency combs or higher accuracy in the line profile measurements is sought for, the combs could be stabilized to a radio-frequency reference, as they often are in frequency metrology, and the adaptive sampling scheme would apply as well. The possibility of using commercial femtosecond lasers or frequency comb systems without any sophisticated optical stabilization scheme considerably ease the implementation of a dual-comb spectrometer and should prompt new applications to real-time spectroscopy and sensing. The adaptive scheme and the averaging [13,14] could also take advantage of digital sampling processors onboard the data acquisition system and would similarly bring suitable real-time processing.

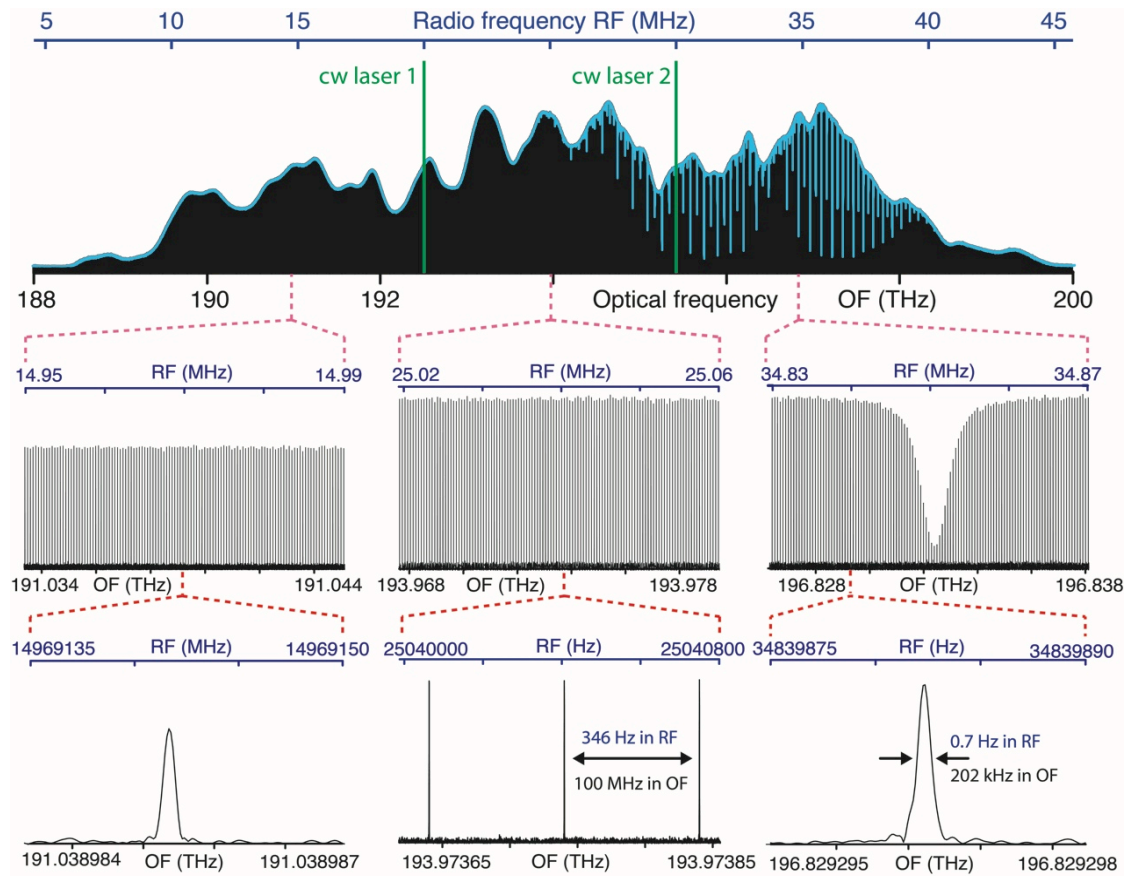

**Figure 5. Adaptive spectrum spanning 12 THz with resolved comb lines.**
The spectrum across the entire domain of emission of the Erbium femtosecond fiber lasers exhibits 120 000 well-resolved individual comb lines. The measurement time is 2.7 s. It reveals several acetylene profiles shaping the discrete comb line intensities. The frequency of the continuous-wave lasers used in the production of the adaptive signals in indicated in green in the upper row of the figure. In the two lowest rows, different degree of zooms in three different spectral regions well apart from these two frequencies isolate the individual comb lines and prove that the coherence between the two combs is maintained over the time of the measurement.

Summarizing, through our demonstration of an accessible and robust technique of dual-comb spectroscopy with free-running femtosecond lasers, the recording of distortion-free spectra with Doppler-limited resolution over the full emission bandwidth of a femtosecond oscillator within a few hundreds of microseconds becomes feasible to scientists and engineers unfamiliar with the sophisticated tools of frequency metrology. The key feature of dual-comb spectroscopy, the very short measurement time, is kept optimal.

**Acknowledgements :**
Critical reading of the manuscript by Arthur Hipke is warmly acknowledged. Research conducted in the scope of the European Laboratory for Frequency Comb Spectroscopy. Support by the Max Planck Foundation, the Munich Center for Advanced Photonics, the E.A.D.S. Foundation, the Conseil Général de l'Essonne, the Triangle de la Physique and the Agence Nationale de la Recherche are gratefully acknowledged.